\documentclass{ws-rv9x6}
\usepackage{subfigure}     
\makeindex
\begin{document}

\chapter {Pairing and realistic shell-model interactions} \label{ra_ch13}

\author{A. Covello}
\address{Istituto Nazionale di Fisica Nucleare, Sezione di Napoli, and
Universit\`a di Napoli Federico II,
Complesso Universitario di Monte S. Angelo,Via Cintia, I-80126 Napoli, Italy}
\author{A. Gargano}
\address{Istituto Nazionale di Fisica Nucleare, Sezione di Napoli} 
\author{T. T. S. Kuo}
\address{Department of Physics, SUNY, Stony Brook, New York 11794, USA}


\begin{abstract}

This paper starts with a brief historical overview of pairing in nuclei, which fulfills the purpose of properly framing the main subject. This concerns the pairing properties of a realistic shell-model effective interaction which has proved very successful in describing nuclei around doubly magic $^{132}$Sn.  We focus attention on the two nuclei $^{134}$Te and $^{134}$Sn with two valence protons and neutrons, respectively. Our study brings out the key role of one particle-one hole excitations in producing a significant difference between  proton and neutron pairing in this region.
\end{abstract}

\body

\section{Introduction and brief historical overview}\label{ra_sec1}

The concept of pairing has been a key concept in nuclear physics 
over the past six decades. As a matter of fact, this concept was the basis
for the famous coupling rules of the Mayer-Jensen shell model 
\cite{Mayer50,Mayer55}. It was not until the end of 1950s, however, that the way to 
a quantitative study of pairing in nuclei was open. 
As is well known, the starting point  was the analogy \cite {Bohr58} between the energy gap in the electronic excitation of superconductors and that existing in the intrinsic excitation spectrum of nuclei. This suggested that the methods developed in the theory of superconductivity in the biennium 1957-1958 \cite{BCS57,Bogoliubov58,Valatin58} could be profitably applied to describe pairing effects in nuclear structure. The first studies along these lines \cite{Belyaev59, Kisslinger60} proved to be quite successful and since then a vast literature has accumulated on the subject of pairing in nuclear systems.  Roughly speaking, this has developed along two directions. On the one side, the focus has been placed on the pairing force itself, including efforts to devise many-body techniques  to go beyond the BCS approximation. On the other side, the pairing properties of both finite nuclei and nuclear matter and  their explanation in terms of effective nuclear interactions have been the main subject of investigation. To put the content of the present paper into its proper perspective, in this Introduction we will briefly touch on both the above lines of research.  
A detailed review of developments in the field through 2003 is given in Ref. \cite{Dean2003}.  

For many years the most widely used approach to the treatment of pairing correlations in nuclei has remained the BCS theory, mainly because of its computational simplicity. Much work, however, has been done to remedy the 
problem of particle-number violation which may be quite serious in finite systems.
In this respect, one may say that this severe drawback of the BCS theory has been a merit on disguise, since it has stimulated the development of several number-conserving approaches to the pairing problem. An historical overview  of the early efforts to improve on the BCS approximation is of course well beyond the scope of this paper. A concise account of the various developments including references through 1985 can be found in Ref. \cite{Andreozzi85},  where an equations-of-motion approach to the pairing problem was proposed. Studies of both even and odd Ni \cite{Covello86} and Sn isotopes \cite{Covello90} evidenced the remarkably good accuracy of this method as compared to that of the standard BCS approximation. In this regard, a word of caution is still in order nowadays when using the latter approximation in nuclear many-body problems. 

While a simple pairing force is able to reproduce remarkably well some nuclear properties (see for instance Refs. \cite{Andreozzi90,Andreozzi92}), it was clear from the beginning that it is inadequate for detailed quantitative studies. In fact, already in the early calculations of Ref. \cite{Kisslinger60} the shell-model effective interaction was taken to be a pairing plus quadrupole force.
In this connection, it is worth noting that in most of the work aimed at improving on the BCS theory
the study of the pairing model has been considered as a first step toward the treatment of more realistic
interactions. This is indeed the case of the equations-of-motion method of Ref. \cite {Andreozzi85}, which
was further developed later \cite{Covello90} to treat a general Hamiltonian within the framework of the seniority scheme. For systems of like nucleons this scheme is particularly appropriate just because of  the  key role played by the pairing component of the nuclear force. 

Since the early 1960s through the mid 1990s a large number of shell-model calculations have been carried out employing a variety of two-body effective interactions. These have gone from  
the so-called ``schematic interactions", like the above mentioned pairing plus quadrupole, which give an oversimplified representation of the real potential, to more complete interactions including operatorial terms consistent with those present in the interaction  between free nucleons.  Clearly, in the latter  the pairing force does not appear explicitely, but manifests its importance through
the $J=0$ matrix elements. One may then speak about ``the pairing properties" of the shell-model effective interaction.  

In the last two decades great progress has been made \cite{Coraggio09} in the derivation of effective interactions from realistic free nucleon-nucleon potentials. These interactions, that contain no free parameters, have been very successful in application providing an accurate description of nuclear structure properties in many cases \cite{Coraggio09}. Based on these results, in the last few years the use of  realistic effective interactions has been rapidly gaining ground in nuclear structure theory. 
In this context, it is clearly of great relevance to investigate their pairing properties and try to find out what is their microscopic origin. The purpose of the present paper is just to discuss this problem. 

We first give a brief sketch of our approach to the derivation of the shell-model effective interaction from the free nucleon-nucleon ($NN$) potential. We then examine the pairing properties of the bare and effective two-body interaction, focusing attention on the role of core polarization in nuclei with like valence nucleons in the $^{132}$Sn  region. 

\section{Outline of theoretical framework}

  As mentioned in the Introduction, this paper is concerned with the pairing properties of shell-model effective interactions derived microscopically from the $NN$ potential. For the sake of completeness,
we give here an outline of the essentials of this derivation. A detailed description including references to original literature can be found in the review paper \cite{Coraggio09}.

The Schr\"odinger equation for a system of A nucleons interacting via two-body forces can be written

\begin{equation}
H\Psi_i=(H_{0}+H_{1})\Psi_i = E_i \Psi_i, \label{Schr} \end{equation}

\noindent
 where
\begin{equation}
H_{0} = T+U
\label{defh0}
\end{equation}
\noindent
and
\begin{equation}
H_{1}= V_{NN}-U,
\label{defh1}
\end{equation}

\noindent
$T$ being the kinetic energy and $U$ an auxiliary  potential introduced to define a convenient single-particle basis.
The effective interaction $V_{\rm eff}$ acting only within a reduced model space is then defined through the eigenvalue problem

\begin{equation}
PH_{\rm eff}P| \Psi_\alpha\rangle = P(H_{0}+V_{\rm eff})P| \Psi_\alpha\rangle= E_\alpha P \Psi_\alpha, \label{defheff} \end{equation}

\noindent
where the  $E_\alpha$  and the corresponding $\Psi_\alpha$ are a subset of the
eigenvalues  and eigenfunctions of the original Hamiltonian defined in the complete Hilbert space. The $P$ operator projects onto the chosen model space, which is defined in terms of the eigenvectors of the unperturbed Hamiltonian $H_{0}$. 

A well-established approach to the derivation of realistic effective interactions from the free $NN$ potential $V_{NN}$ is provided by the $\hat Q$-box folded-diagram expansion \cite{Coraggio09}.  
In our calculations, we have used as initial input the CD-Bonn potential \cite{Machleidt01}. The existence of a strong repulsive core, however, makes this potential unsuitable for perturbative calculations and hence requires a renormalization procedure. This we do through use of the $V_{\rm low-k}$ approach \cite{Bogner02}, which has proved to be an advantageous alternative to the traditional Brueckner $G$-matrix method. More precisely, we construct a 
smooth low-momentum potential, $V_{\rm low-k}$, by integrating out the high-momentum modes of 
$V_{NN}$ down to a cutoff momentum $\Lambda$. This integration is carried out with the requirement that the deuteron binding energy and phase shifts of $V_{NN}$ up to $\Lambda$ are preserved by $V_{\rm low-k}$. 

Once the $V_{\rm low-k}$ is obtained, we use it, plus the Coulomb force for protons, as input interaction for the derivation of $V_{\rm eff}$. The calculation of the $\hat{Q}$-box , which is a sum of irreducible valence-linked diagrams, is performed at second order in the interaction. That is to say, we include four two body terms: the $V_{\rm low-k}$, the two core polarization diagrams $V_{\rm 1p1h}$ and $V_{\rm 2p2h}$, corresponding to one particle-one hole and two particle-two hole excitations, and a ladder diagram accounting for excluded configurations above the chosen model space. The shell-model effective interaction is finally obtained by summing up the $\hat Q$-box folded diagram series using the Lee-Suzuki iteration method \cite {Lee80}. 

In this paper, we present results of realistic shell-model calculations wherein doubly magic $^{132}$Sn
is assumed to be a closed core. We let the valence protons occupy the five levels $0g_{7/2}$, 
$1d_{5/2}$, $1d_{3/2}$, $2s_{1/2}$, and $0h_{11/2}$ of the 50-82 shell, while for neutrons  the model space includes the six levels $0h_{9/2}$, $1f_{7/2}$, $1f_{5/2}$, $2p_{3/2}$, $2p_{1/2}$, and  $0i_{13/2}$ of the 82-126 shell. As regards the choice of single-proton and -neutron energies, they have been taken from experiment. The adopted values are reported in \cite{Coraggio09b} and  \cite{Covello11} for protons and neutrons, respectively. As mentioned above, the two-body effective interaction has been derived from the CD-Bonn potential renormalized through  the  $V_{\rm low-k}$ procedure. Details on the derivation can be found in \cite{Covello11}.

Before closing this section, one more remark is in order. 
As is well known, there is no unique $V_{NN}$. Rather, there are several high-quality potentials which fit equally well ($\chi^2/{\rm datum} \approx 1$) the $NN$ scattering data up to the inelastic threshold, namely they are phase-shift equivalent.  This may raise the question of how much nuclear structure results may depend on the choice of the $NN$ potential one starts with. 
A detailed description of the derivation of $V_{\rm low-k}$ from $V_{NN}$ as well as a discussion of its main features can be found in \cite{Coraggio09,Bogner02}.
We only want to point out here that the use of $V_{\rm low-k}$ largely reduces the ambiguity in the choice of $V_{NN}$. In fact, we have verified \cite{Coraggio09} that shell-model effective interactions derived from phase-shift equivalent $NN$ potentials through the $V_{\rm low-k}$ approach lead to very similar results. In other words, $V_{\rm low-k}$ gives an approximately unique representation of the $NN$ potential.

\section {Realistic effective interaction and pairing properties in the $^{132}$Sn region: the two valence-particle nuclei $^{134}$Te and $^{134}$Sn}

We have investigated the pairing properties of a realistic effective interaction constructed for the
$^{132}$Sn region, focusing on the two nuclei $^{134}$Te and $^{134}$Sn. These nuclei, with two valence protons and neutrons, respectively, allow a direct study of the main features of the proton-proton and neutron-neutron interaction in the presence of the same closed core. This is particularly interesting because the available experimental data have evidenced a large difference between proton and neutron energy gaps. More precisely, while the first $2^+$ state in $^{134}$Te lies at 1.28 MeV, the excitation energy of this state in $^{134}$Sn drops to 726 keV, making it the lowest first-excited $2^+$ level observed in a semi-magic even-even nucleus over the whole chart of nuclides. 

This, as well as other peculiar features (see for instance Ref. \cite{Covello07}) of nuclear structure in the $^{132}$Sn region for $N > 82$,  has tended to support the idea that
the neutron excess produces a quenching of the $N=82$ shell closure, in contrast with the 
interpretation \cite{Terasaki02,Shimizu04} based on a reduction  of the neutron pairing above the $N=82$ shell.
Evidence for a shell quenching was also considered a first mass measurement of $^{134}$Sn \cite{Fogelberg99}, which casted doubt on the doubly magic nature of
$^{132}$Sn. However, a high-precision Penning trap mass measurement \cite{Dworschak08} has revealed a 0.5 MeV discrepancy with respect to the previous measurement, which restores the neutron-shell gap at $N=82$.

Nuclei around $^{132}$Sn, below and above $N=82$, have been the subject of several realistic shell-model calculations (\cite{Coraggio09} and references therein, \cite{Covello11,Covello10}) which have yielded very good agreement with experiment without invoking any shell-structure modification. 
In particular, the properties of $^{134}$Te, which exhibits a ``normal" proton pairing, as well as those of $^{134}$Sn with a weak neutron pairing, are well described by our realistic effective interaction, as is seen in Fig. 1, where the experimental energies of the first three excited levels are compared with the calculated ones. Especially worthy of note is that the energy of the $2^+$ state in both nuclei, namely the proton and neutron gap, is remarkably well reproduced. This clearly means that our effective interaction possesses good pairing properties.  
\vspace{-.3cm}

\begin{figure}[h]
\vspace{0.1cm}
\begin{center}
\begin{tabular}{lr}
\hspace{-.3cm}
     \resizebox{55mm}{!}{\includegraphics{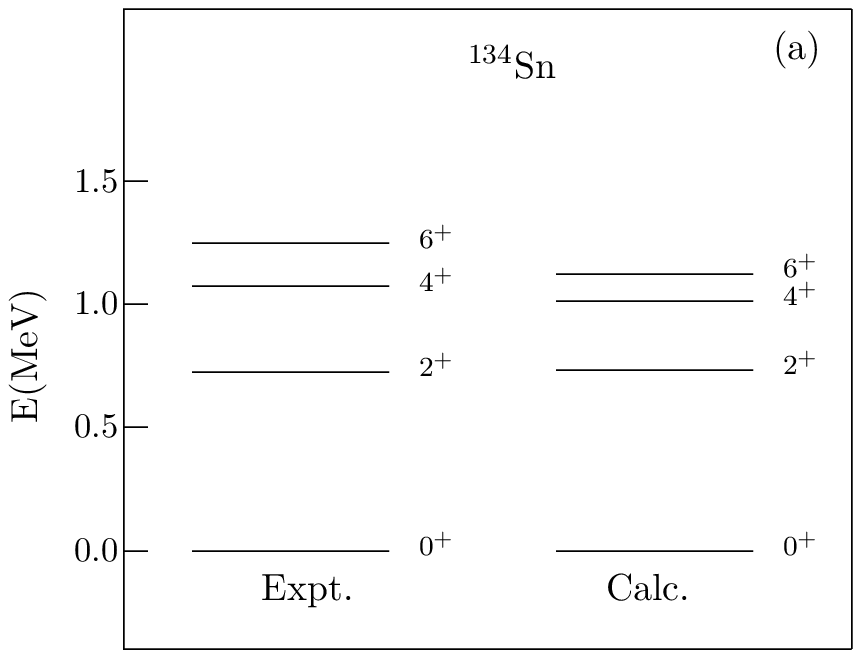}}
\hspace{+0.1cm}
     \resizebox{55mm}{!}{\includegraphics{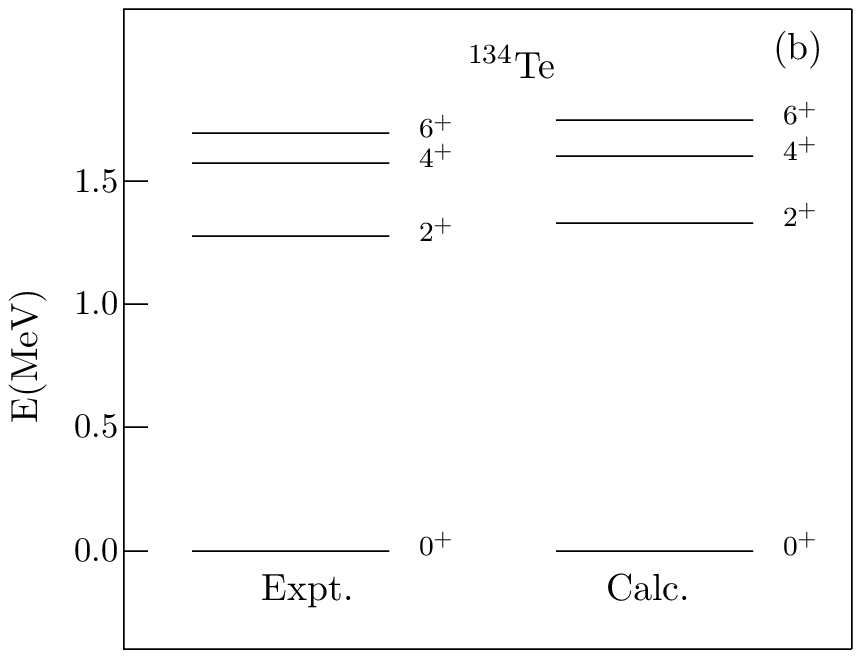}} \\
\end{tabular}
\end{center}
\vspace{-0.4cm}
\caption{ Experimental and calculated spectrum of $^{134}$Sn (a) and $^{134}$Te (b).}

\end{figure}
\vspace{-.3cm}

As discussed in the previous section, we know, however, that a realistic effective interaction is obtained from the nucleon-nucleon potential through a perturbation procedure which takes 
microscopically into account core polarization effects and excluded configurations above the model space.  It is therefore highly interesting to investigate the mechanism that lies behind its pairing properties.

In the following we shall try to go deeper into this subject.  However, before doing this two comments are in order. First, all states  reported in Figs. 1 have a weak configuration mixing. More precisely, the ground state as well as the three excited states  are dominated by the $(\pi g_{7/2})^2$ configuration in $^{134}$Te while by the $(\nu f_{7/2})^2$ configuration in  $^{134}$Sn. This makes it possible to only focus  our attention on the matrix elements involving these two configurations.

The second comment concerns the Coulomb energy. As mentioned in Sec. 2, our effective interaction includes the Coulomb force which is added directly to $V_{\rm low-k}$ before calculating the diagrams composing the $\hat Q$-box. Since we are interested in the pairing properties of the nucleon-nucleon force in the nuclear medium, from now on we shall consider an effective interaction derived without inclusion of the Coulomb force. We only mention here that the Coulomb force affects the matrix elements of the effective interaction we are interested in  by a quantity that ranges from about 150 to 250 keV. The largest contributions come from the bare Coulomb force, namely from first-order diagrams, while second-order and folded diagrams are generally quite small.

\vspace{-.5cm}

\begin{figure}[h]
\vspace{0.1cm}
\begin{center}
\begin{tabular}{lr}
\hspace{-.4cm}
     \resizebox{57mm}{!}{\includegraphics{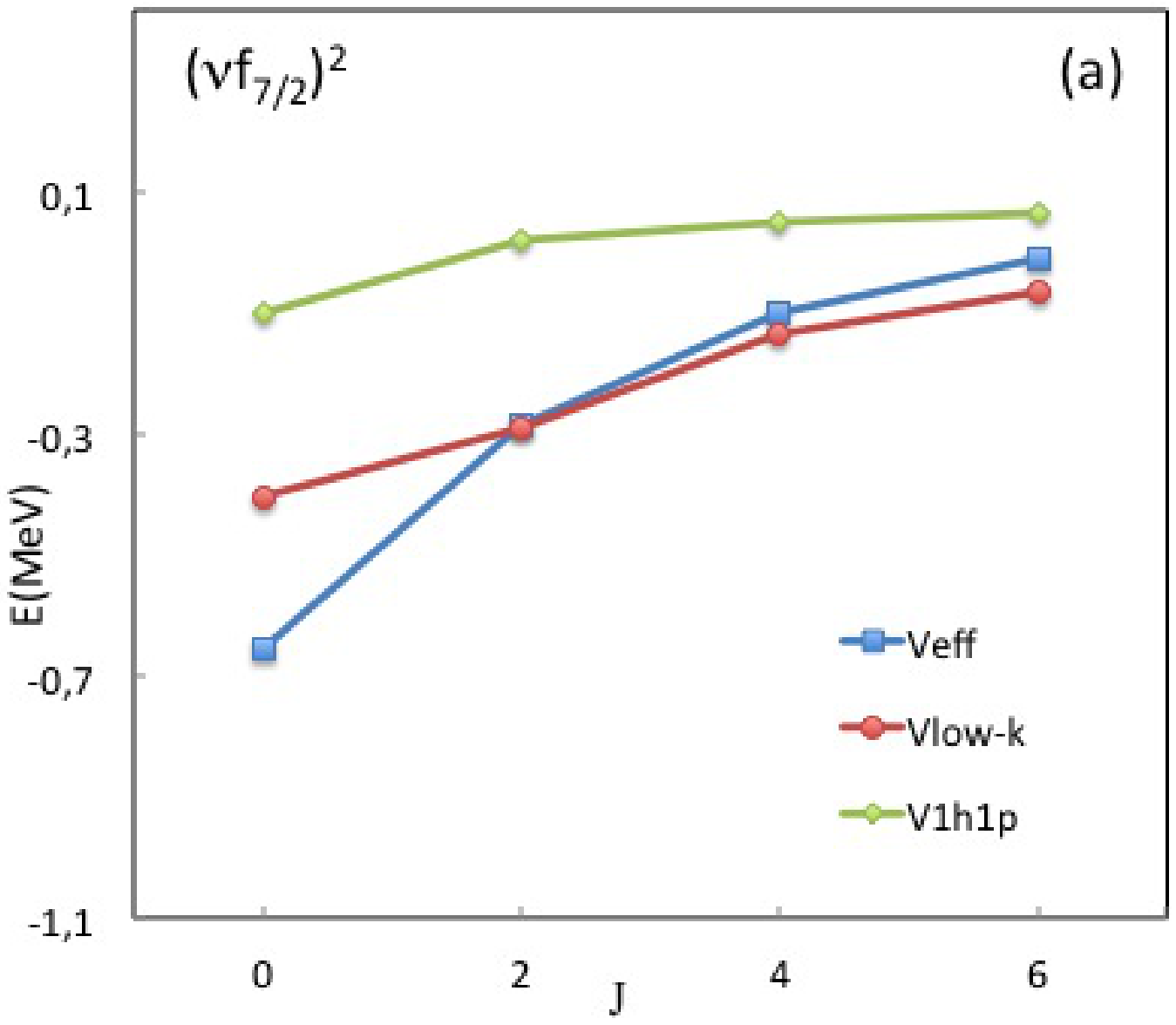}}
     \resizebox{57mm}{!}{\includegraphics{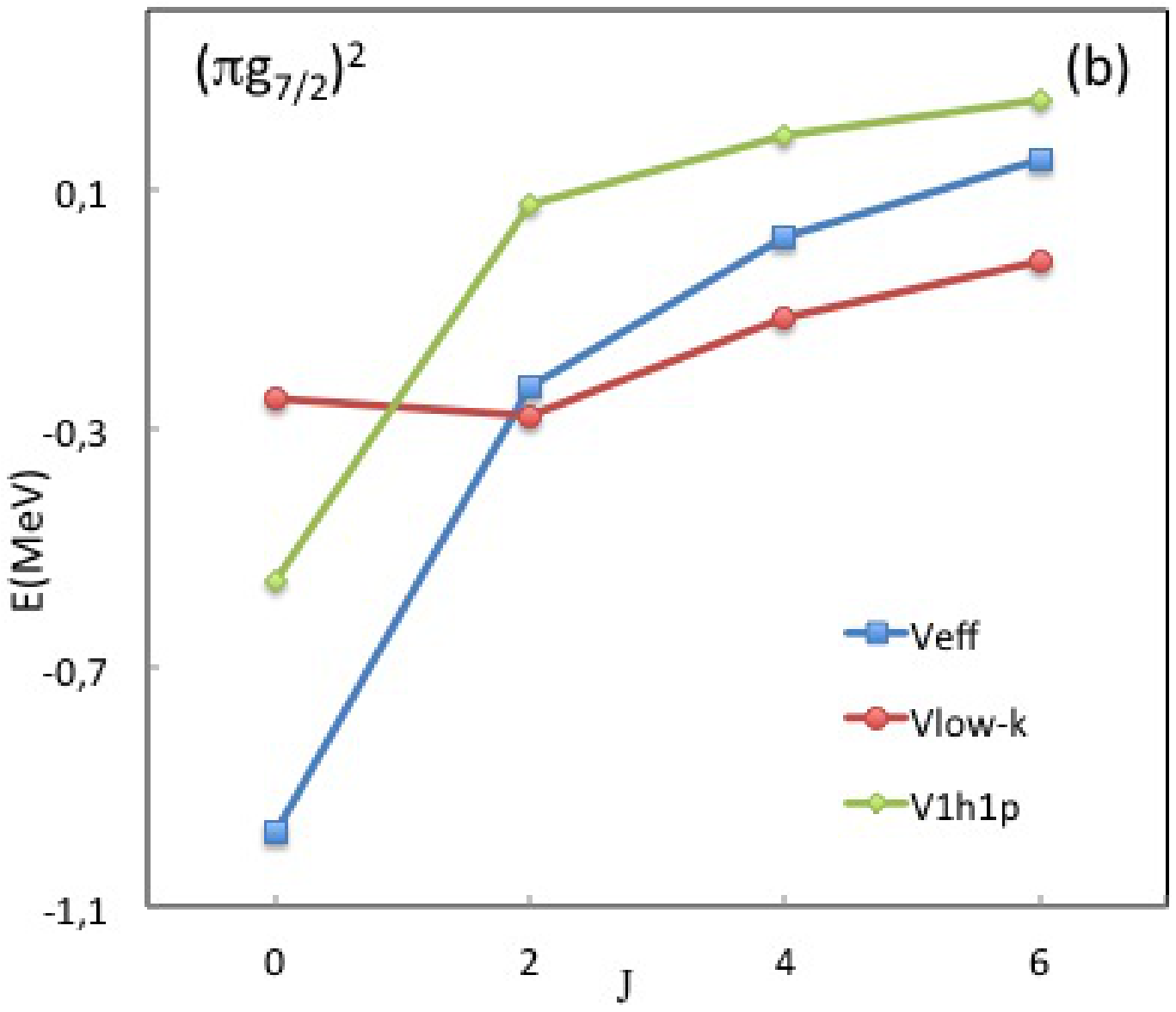}} \\
\end{tabular}
\end{center}
\vspace{-0.9cm}
\caption{ Matrix elements of $V_{\rm eff}$, $V_{\rm low-k}$, and $V_{\rm 1p1h}$ for the 
$(\nu f_{7/2})^2$ (a) and $(\pi g_{7/2})^2$ (b) configuration. }
\end{figure}
\vspace{-0.4cm}

In Fig. 2 we show the matrix elements of  $V_{\rm eff}$, $V_{\rm low-k}$, and $V_{\rm 1p1h}$ for the  $(\nu f_{7/2})^2$  and $(\pi g_{7/2})^2$ configurations. Note that in the analysis of  $V_{\rm eff}$ we consider only 1p1h contributions, because the other diagrams play a minor role in our discussion.
From Fig. 2(a) we see that the behavior of $V_{\rm eff}$ as a function of the angular momentum is similar to that of the bare $V_{\rm low-k}$ interaction, the latter being only
slightly modified by the 1p1h contributions. The $V_{\rm 1p1h}$ curve is in fact almost flat around 0 MeV, showing a non negligible decrease only for $J^{\pi}=0^+$.
Actually, the $J^{\pi}=0^+$ matrix element  of $V_{\rm low-k}$, which is the most affected one by medium effects, is shifted down by only 250 keV including all contributions.
In practice, the behavior exhibited by the $(\nu f_{7/2})^2$ multiplet is a direct manifestation of  the $V_{\rm low-k}$ interaction, which is primarily responsible for the weakness of the neutron-neutron pairing component.
  
The situation is just reversed when we look at the curves of Fig. 2(b) for the $(\pi g_{7/2})^2$ configuration, which evidence  the crucial role of the  1p1h core polarization in determining the proton-proton pairing.  In other words, $V_{\rm low-k}$ does not possess the right pairing properties, the $J^{\pi}=0^+$ and $2^+$ matrix elements being almost  equal and quite close to  those with $J^{\pi}=4^+$ and $6^+$. On the contrary,  $V_{\rm 1p1h}$ has  a  $J^{\pi}=0^+$ matrix element much larger than those with $J \ne 0$, which explains the gap existing between the ground and the $2^+$ states in the spectrum of  $^{134}$Te.

In summary, the effective pairing interaction between two protons and two neutrons outside $^{132}$Sn  may be mainly traced to different core renormalizations of the $V_{\rm low-k}$ $NN$ potential. For protons, the dominant role is played by the one particle-one hole excitations, which produces a sizeable energy gap. These core excitations play instead a minor role for neutrons, which results in a reduced  neutron pairing. 
 
\vspace{-.4cm}

\begin{figure}[h]
\vspace{0.1cm}
\begin{center}
\begin{tabular}{lr}
\hspace{-.3cm}
     \resizebox{55mm}{!}{\includegraphics{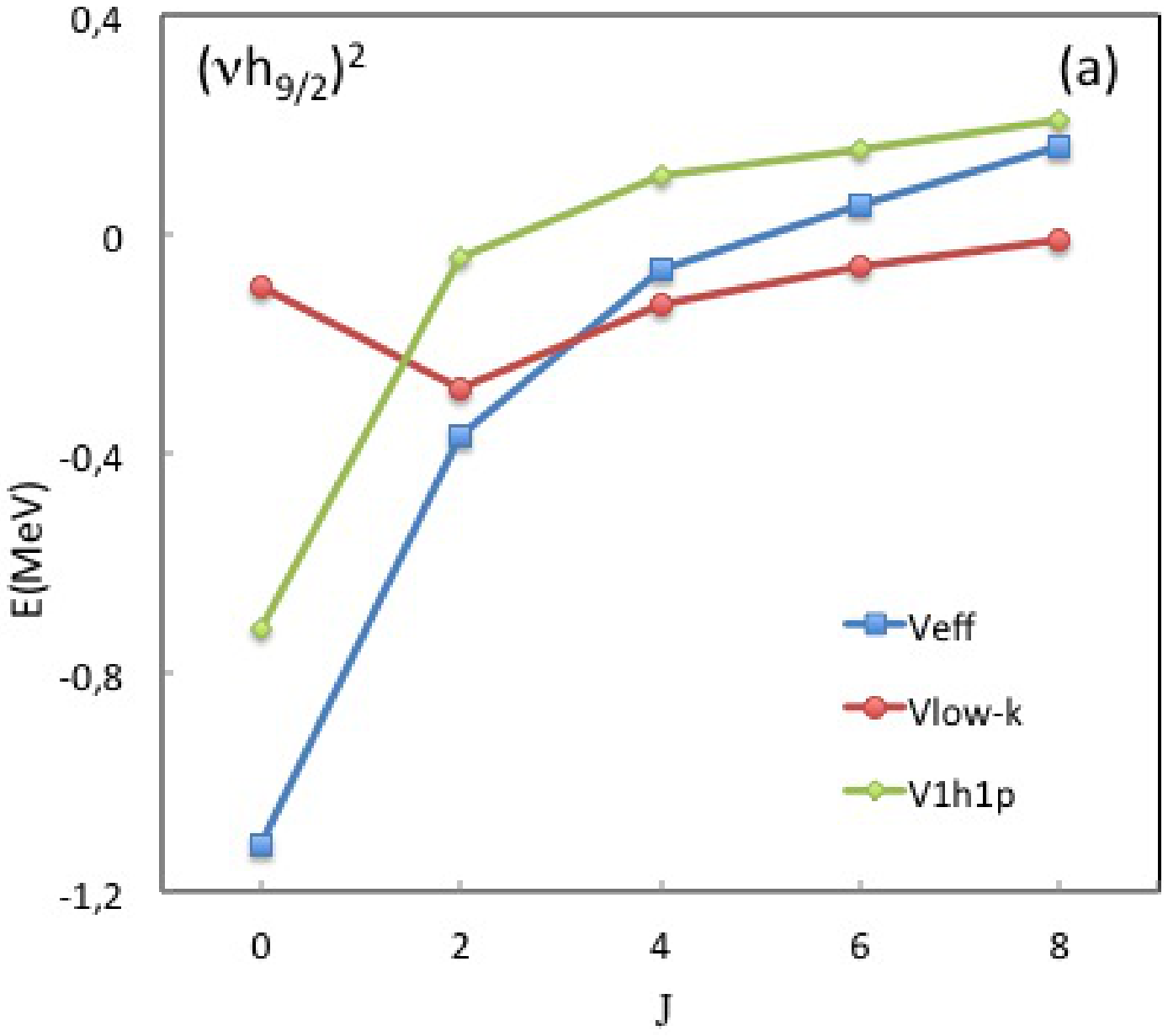}}
\hspace{+0.1cm}
     \resizebox{55mm}{!}{\includegraphics{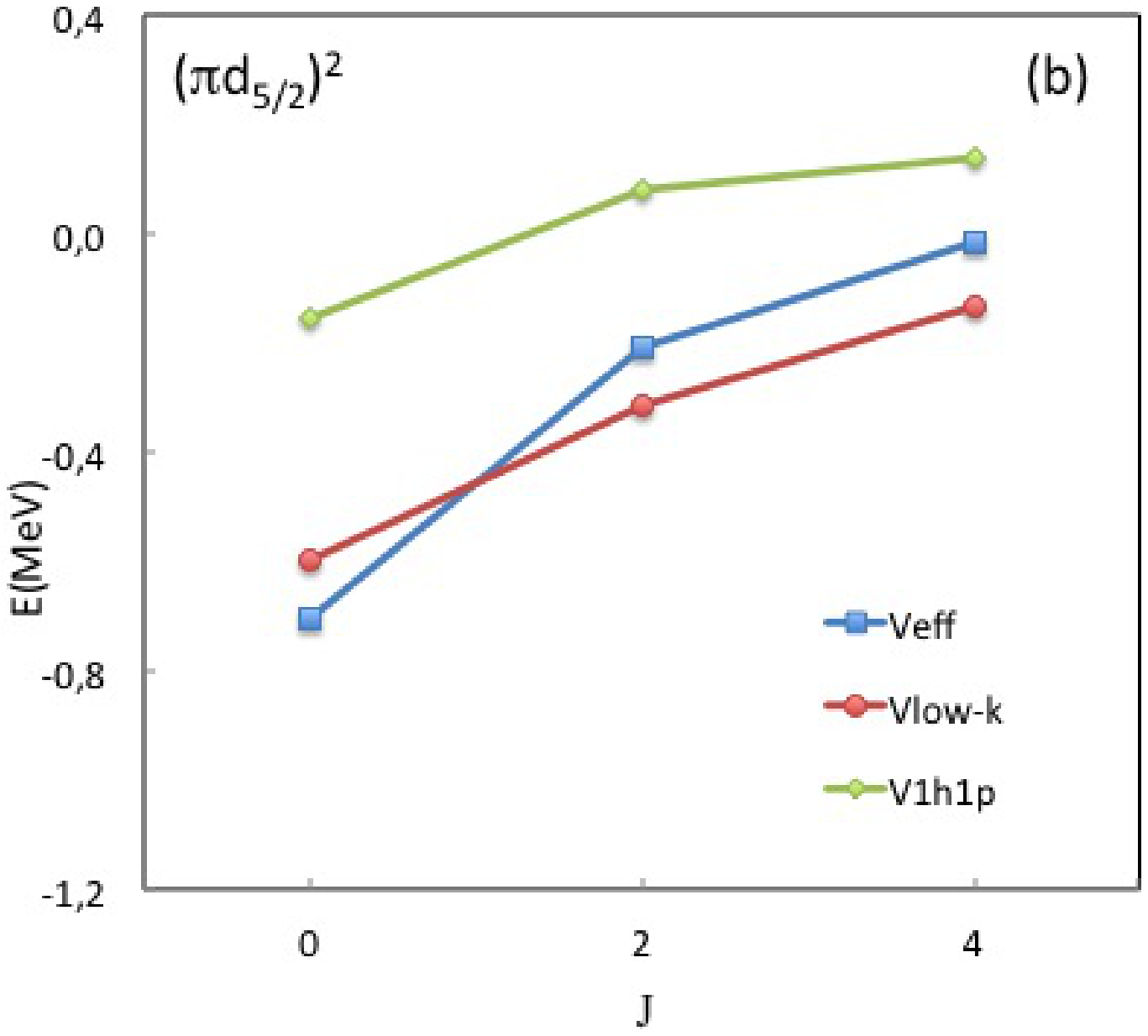}} \\
\end{tabular}
\end{center}
\vspace{-0.7cm}
\caption { Matrix elements of $V_{\rm eff}$, $V_{\rm low-k}$, and $V_{\rm 1p1h}$ for the 
$(\nu h_{9/2})^2$ (a) and  $(\pi d_{5/2})^2$ (b) configuration.}
\end{figure}

\vspace{-.3cm}

In the above context, it is worth noting that the proton $g_{7/2}$ level has its spin-orbit partner in the lower 28-50 shell, while this is not the case for the neutron $f_{7/2}$ level,
whose spin-orbit partner $f_{5/2}$ lies in the same 82-126 shell. Based on this fact, we have found it interesting to perform the same analysis for different proton and neutron orbits.
To this end, we have considered the $d_{5/2}$ level for protons and the $h_{9/2}$ for neutrons, the latter having its spin-orbit partner in the lower shell. The matrix elements of $V_{\rm eff}$, $V_{\rm low-k}$, and $V_{\rm 1p1h}$ for the $(\nu h_{9/2})^2$ and $(\pi d_{5/2})^2$  configurations are reported in Fig. 3.

We see that the protons have now a weak pairing while the neutrons receive a large contribution from 
$V_{\rm 1p1h}$. This confirms the crucial role of core polarization in determining the pairing properties 
of the shell-model effective interaction, especially when the two spin-orbit partners are in different shells. In this regard, suffice it to note that the $J=0$ matrix elements of  $V_{\rm 1p1h}$ for the configurations $(\pi g_{7/2})^2$ and 
$(\nu h_{9/2})^2$ are, respectively,  more than twice and more than half an order of magnitude larger than those of $V_{\rm low-k}$.

\vspace{-.2cm}

\begin{table}[ht]
\tbl{$<j^{2}; J^{\pi}=0^{+}|V|j^{2}; J^{\pi}=0^{+}>$ neutron 
matrix  elements (in MeV) for  $V_{\rm eff}$,  $V_{\rm low-k}$, and  $V_{\rm 1p1h}$. 
 $G$ stands for the
pairing constant (see text for definition).}
{\begin{tabular}{@{}lcccc@{}} \toprule
$j$ & $V_{\rm eff}$ & $V_{\rm low-k}$& $V_{\rm 1p1h}$ & $G$  \\
\colrule
$f_{7/2}$ & -0.654 & -0.403 & -0.100 & -0.16 \\
$p_{3/2}$ & -0.524& -0.482& 0.008 & -0.26\\
$h_{9/2}$ & -1.114 & -0.098 & -0.723 & -0.22\\
$p_{1/2}$ & -0.179&  0.003& -0.104 & -0.18\\
$f_{5/2}$ & -0.404 & -0.101& -0.117& -0.13\\
$i_{13/2}$ & -0.974& -0.187& -0.464& -0.14 \\
\colrule
\end{tabular}
}
\label{ra_tbl2}
\end{table}

\vspace{-.4cm}

\begin{table}[ht]
\tbl{$<j^{2}; J^{\pi}=0^{+}|V|j^{2}; J^{\pi}=0^{+}>$ proton matrix  
elements (in MeV) for  
$V_{\rm eff}$,  $V_{\rm low-k}$, and  $V_{\rm 1p1h}$. $G$  stands for the
pairing constant (see text for definition).}
{\begin{tabular}{@{}lcccc@{}} \toprule
$j$  & $V_{\rm eff}$ & $V_{\rm low-k}$& $V_{\rm 1p1h}$ & $G$  \\
\colrule
$g_{7/2}$ & -0.967&  -0.248& -0.554 & -0.24 \\
$d_{5/2}$ & -0.705&  -0.600& -0.156 & -0.23 \\
$d_{3/2}$ & -0.370&  -0.162& -0.134 & -0.18 \\
$s_{1/2}$ & -0.560&  -0.860&  0.181 & -0.56\\
$h_{11/2}$ & -1.143&  -0.379& -0.759 & -0.19\\
\colrule
\end{tabular}
}
\label{ra_tbl3}
\end{table} 

\vspace{-.2cm}

To conclude this section, we focus attention on the $J=0$ matrix elements to see how they compare with the coupling strength of a pairing force. 
In Tables 1 and 2 the $J=0$ diagonal matrix
elements of $V_{\rm eff}$, $V_{\rm low-k}$,  and $V_{\rm 1p1h}$ are reported for the 6 neutron
orbits of the 82-126 shell and the 5 proton orbits of the 50-82 shell. 
The last column of both tables shows 
$G= <j^{2}; J^{\pi}=0^{+}|V_{\rm eff}|j^{2}; J^{\pi}=0^{+}>/(j+1/2)$.
We see that the proton pairing constant is on the overall stronger than the neutron one. In fact, the  values of $G$ for protons range from 0.18 to 
0.24 MeV without considering the 0.56 MeV peak for the $s_{1/2}$ orbit, while for neutrons from 0.13 to 
0.26 MeV. Actually, most of these values are not far from that obtained from the relation  $G \simeq 20/A$, and  lie within  the limits deduced from empirical analyses of data in the $A=132$ mass region \cite{Andreozzi90}. 
Finally, as  regards the contribution of the 1p1h core polarization, it is worth noting  that the  comments made for the configurations considered above also apply to the other configurations reported in Tables 1 and 2.

\section{Summary and concluding remarks}

In this paper, we have discussed the role of pairing in nuclei as it is determined by
a shell-model effective interaction derived from the free $NN$ potential 
without use of any adjustable parameter. 
To this end, we have focused attention on the two nuclei $^{134}$Te and $^{134}$Sn which have
respectively two protons and two neutrons outside doubly magic $^{132}$Sn. These nuclei 
are currently a subject of great interest. A peculiar feature, which has been a matter of debate, is certainly the large difference between the proton and neutron energy gap, the latter being about 0.5 MeV smaller than the former. 

Our effective interaction has been derived within the framework of perturbation theory starting from the CD Bonn $NN$ potential, whose strong short-range repulsion has been renormalized by means of the $V_{\rm low-k}$ approach. This accounts remarkably well for the  reduction in the neutron pairing gap without any need of shell-structure modifications.   

To try to understand the microscopic origin of the pairing component of the $NN$ potential in the nuclear medium, we have made a detailed analysis of $V_{\rm eff}$. As a main result, we have evidenced the essential contribution of the 1p1h core polarization to its pairing properties.

As recalled in the Introduction, pairing is  a key concept in nuclear physics since a long time.
It is very gratifying that the great progress made in constructing realistic effective interactions for shell-model calculations makes now possible microscopic studies of pairing properties.

\end{document}